\documentstyle[aps,prb,twocolumn,psfig,epsfig,overcite]{revtex}

\def\mue{\mu_{\rm eff}}
\def\epse{\varepsilon_{\rm eff}}
\def\epsb{\varepsilon_{\rm Board}}
\def\epsm{\varepsilon_{\rm m}}
\def\nne{n_{\rm eff}}

\def\be{\begin{equation}}
\def\ee{\end{equation}}

\begin{document}

\title{Transmission Losses in Left-handed Materials}

\author{P. Marko\v{s}$^*$, I. Rousochatzakis and C. M. Soukoulis\\
Ames Laboratory and Department of Physics and Astronomy,\\ 
Iowa State University, Ames, Iowa, 50011}

\maketitle

\begin{abstract}
We numerically analyze  the origin of the transmission 
losses in  left-handed
structures. Our  data confirms that left handed structures
can have very good  transmission properties,  in spite of the expectable
dispersion of their {\sl effective} permeability  
and refraction index. 
The large  permittivity of the metallic components  improves the transmission.
High losses, observed in recent experiments, could be explained by
the absorption of  the dielectric board.

\end{abstract}

\vspace*{1cm}

\noindent PACS numbers: 41.20.Jb, 42.25.Bs, 73.20.Mf

\vspace*{1cm}

Left-handed (LH) materials is a common name for the man-made structures 
which posses, in a  given frequency region, both negative 
{\sl effective} electrical permittivity and magnetic permeability. 
Although such materials are in general not available  in nature,
their  experimental fabrication  became possible after   
the suggestion of Pendry {\sl et al}.
\cite{Pendry_1}.  
They predicted, that a lattice of metallic split rings resonators (SRR)
may exhibit, in a resonance frequency region, 
negative {\sl effective} permeability $\mue$. 
It is also well known that a periodic lattice of thin metallic wires
behaves as an effective medium with negative {\sl effective}
permittivity $\epse$
\cite{Soukoulis}.
By combining a lattice of metallic wires with a lattice of SRRs,
Smith {\sl et al}.
\cite{Smith}
created for the first time left handed structures.

At present, LH materials attract a growing interest of both theoretical
and experimental research. Various interesting physical properties
of LH structures were discussed in Refs.
\cite{Veselago} and \cite{Pendry_2}. Pendry 
\cite{Pendry_3} 
suggested  that LH materials enable the construction 
of perfect lens. Smith et al.
\cite{SSMS}
proved, on the basis of the numerical data,
that the LH structure indeed possesses negative refraction
index. The negative refraction of the electro-magnetic (EM) waves 
was experimentally observed in Ref.
\cite{Shelby}.  These unusual results \cite{Smith,Veselago,Pendry_2,Pendry_3,SSMS} have raised
objections both to the interpretation of the experimental data and to the
realization of negative refraction
\cite{Garcia_1,Walser,Garcia_2}.

In spite of the considerable progress in the studies of the LH
materials, a lot of  questions  remained unanswered. 
One of the  most important question
is, whether the LH structures  have propagating solutions.
LH systems must be  dispersive
\cite{Veselago}.
The frequency dependence of 
the {\sl effective} permittivity and permeability of the LH
materials  is
\cite{Pendry_1} 
\be\label{epseff}
\epse(f)=1-\frac{f_e^2}{f^2+i~f\gamma}
\ee
and
\be\label{mueff}
\mue(f)=1- \frac{f_m^2-f_{m0}^2}{f^2-f_{m0}^2+i~f\gamma}.
\ee
In Eqs. \ref{epseff} and \ref{mueff}, $f_e$ ($f_m$) is the electronic (magnetic)
plasma frequency, respectively, $f_{m0}$ is magnetic
resonance frequency,   and $\gamma$ represents 
the losses of the system.
Due to the strong dispersion in the resonance interval,
the  absorption is assumed to be large.
\cite{Landau}
The first experiments
\cite{Smith,Shelby}
indeed reported that the transmission of the LH samples was only -20 dB. 
Recent theoretical analysis 
\cite{Garcia_1} 
of the experimental data  
led even to the conclusion, that the transmission should decrease
exponentially with the thickness of the LH structure.
Contrary to skeptics
\cite{Garcia_1} 
we show in the present  paper that LH structures posses very high
transmittance. 
Our recent numerical simulations
\cite{PRE} already showed  that the transmission of a  LH system
could be as good as for a right-handed system.

\smallskip

\begin{figure}[t]
\epsfig{file=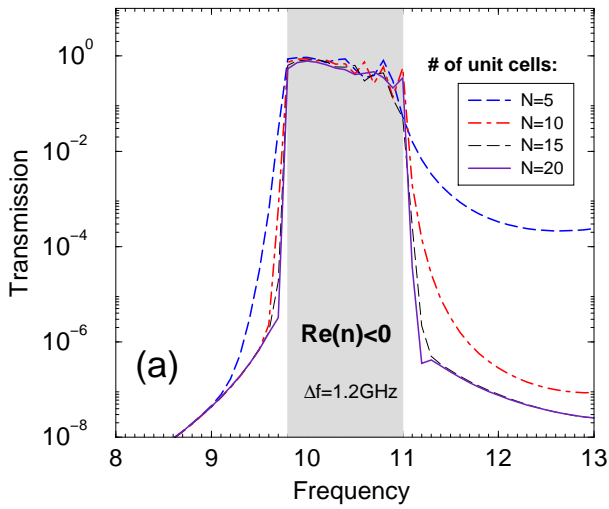,width=8cm}
\begin{center}\epsfig{file=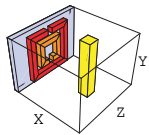,width=3.5cm}
\end{center}


\epsfig{file=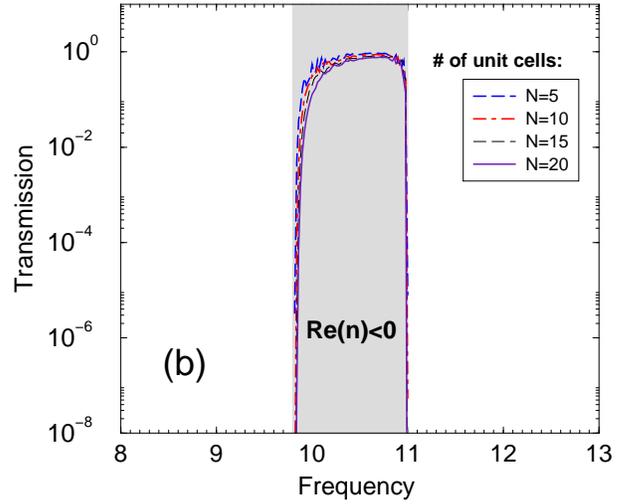,width=8cm}
\caption{Transmission $T$ of the EM wave through the LH structure of
various lengths along the propagation direction. 
(a) Result of transfer matrix simulations. Inset shows the structure
of the unit cell. 
The size of the unit cell is $3.3\times 3.67\times 3.67$ mm.
The simulated system consists of a regular three dimensional 
array of unit cells, infinite in $x$ and $y$ directions.  
EM wave propagates along the $z$ direction. 
(b) transmission of a homogeneous LH slab
with $\epse$ and $\mue$ given by Eqs. \ref{epseff} and \ref{mueff} with 
parameters $f_{m0}=9.8$, $f_m=11$, $f_e=12.8$ and $\gamma=6\times 10^{-5}$ (all frequencies are given in GHz). }
\label{0xL}
\end{figure}

To analyze the transmission properties of LH structures 
in more detail, we first study the
system length dependence of the transmission $T$ for the LH structure
with a  metallic permittivity $\epsm=10^5\times (-3+i~5.88)$.
We will present below also the frequency dependence of $T$ for different values of $\epsm$. 
Fig. \ref{0xL}a  shows the frequency dependence of the transmission for
various system sizes. 
This data was obtained by the use of the transfer matrix (TM) technique
\cite{PRE}.
The simulated
structure was described in details in Ref. \cite{PRE} and is shown as
an inset in fig. \ref{0xL}a.
A resonance interval of $9.8\le f\le 11$ (in GHz),
 in which transmission is close to one, is  clearly visible. 
Fig. \ref{0xL}b  shows the transmission peak for  a  homogeneous
LH model with an effective permittivity and permeability given by Eqs.
\ref{epseff} and \ref{mueff}, respectively. In Eqs. \ref{epseff} and \ref{mueff} we choose
parameters which fit our numerical data, shown in fig. \ref{0xL}a. 
Note that the value of 
$\gamma=6\times 10^{-5}$ GHz
is  three  order of magnitude smaller than that used in Ref. \cite{Shelby}
to interpret the experimental data. This means that there are almost no
loses in our structure. \cite{note_3}

\begin{figure}[t]
\epsfig{file=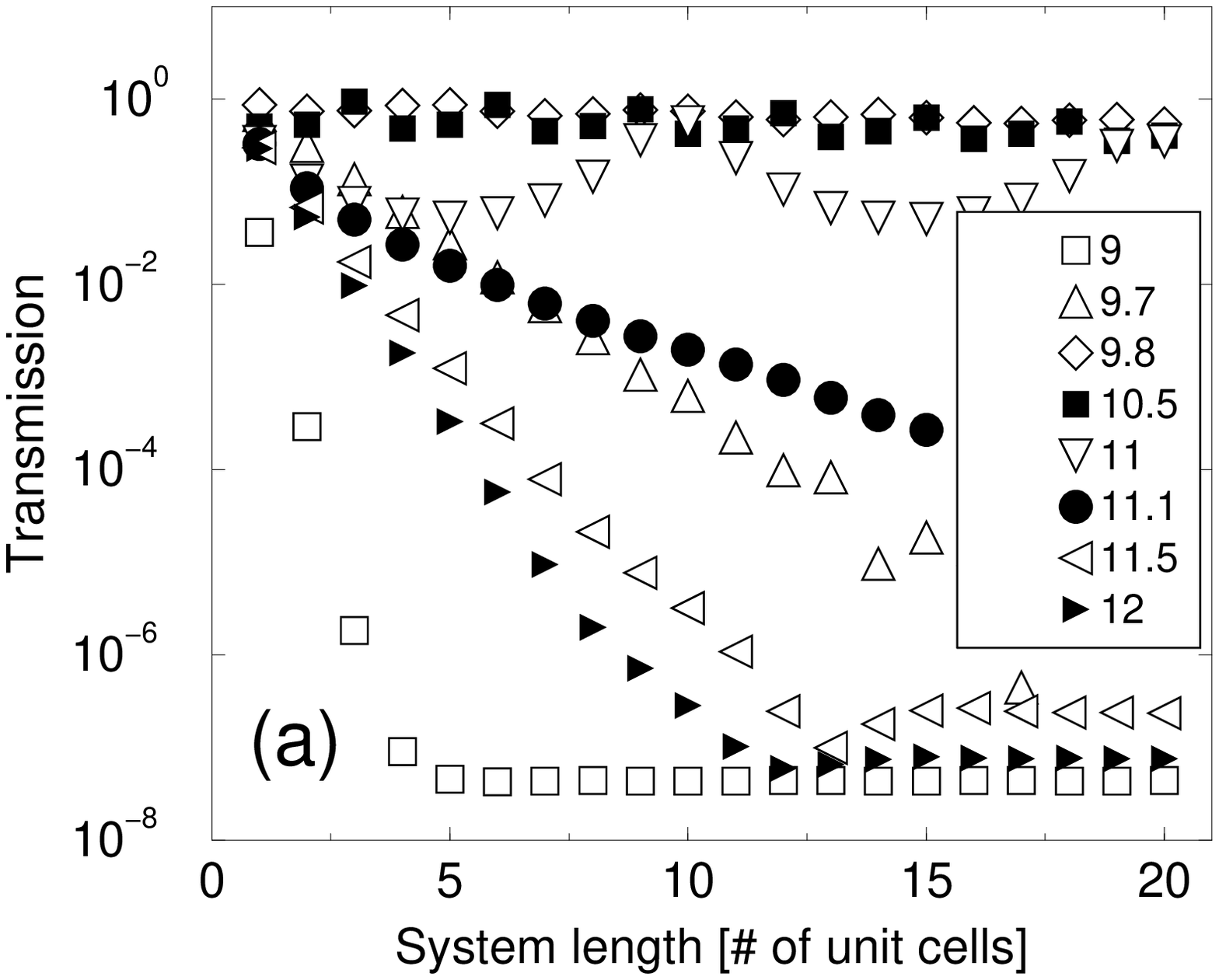,width=8cm}\\
\epsfig{file=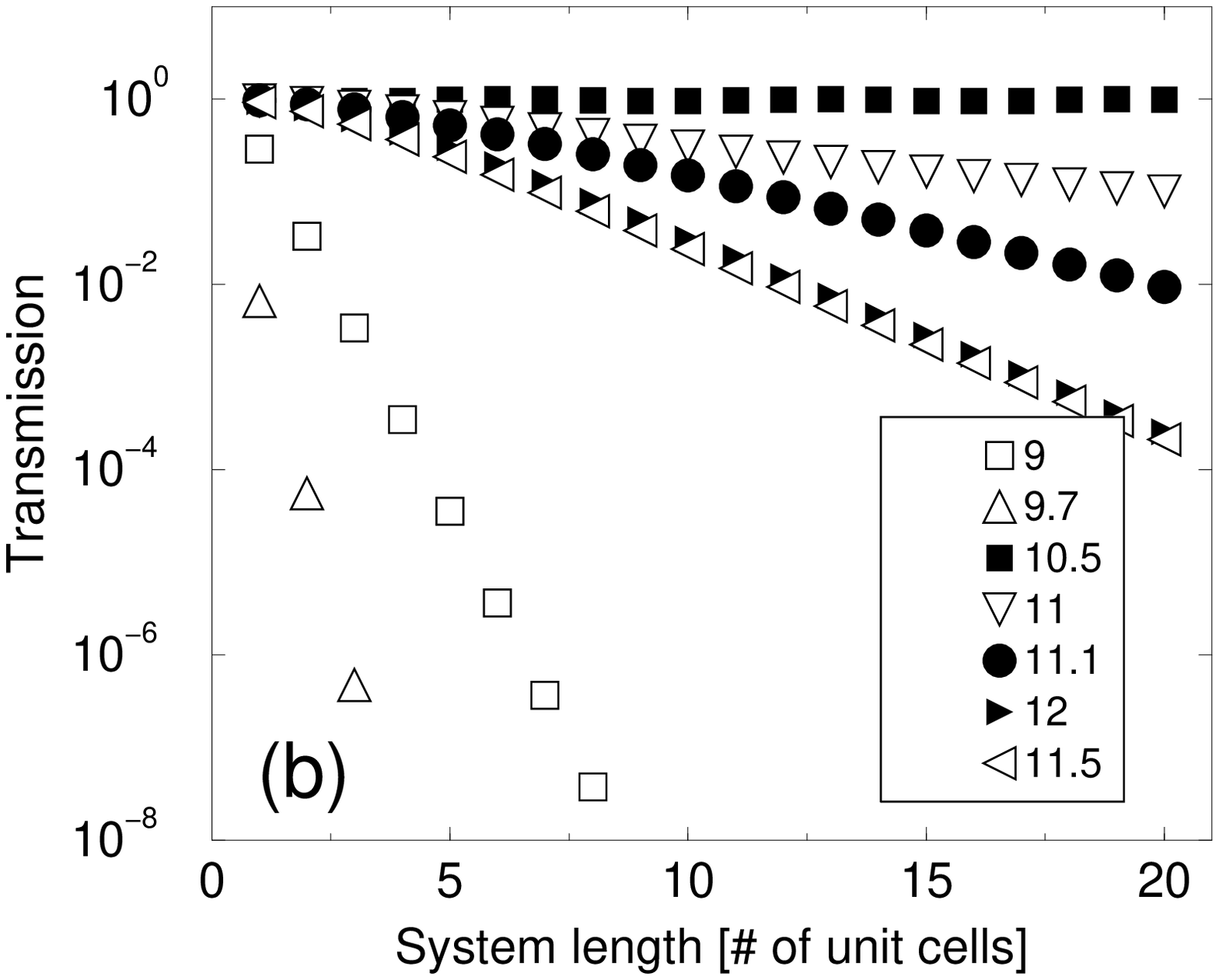,width=8cm}\\
\caption{System length dependence of the transmission $T$ of the EM waves for
 various frequencies. Note that transmission never decreases bellow
 a certain limit.
This is because of the anisotropy of the system. \cite{PRE,unpubl}. 
(a) data from transfer matrix  simulations, (b) data for the  homogeneous model
 with effective permittivity and permeability given by Eqs. \ref{epseff}
and \ref{mueff}
 with parameters listed in fig. \ref{0xL}.
}
\label{0xt}
\end{figure}

In fig. \ref{0xt} we plot the transmission as a 
function of the system length for different frequencies $f$. 
The transmission decreases exponentially with the
system length, when $f$ lies outside the resonance interval.
However, 
for EM waves with frequencies within the resonance interval
only  small decrease of the transmission is observed. 
This unambiguously shows that the transmission is really high
in LH materials with realistic parameters for the
permittivity of metal. This is correct despite
the fact that Im $\epsm$ is of the order of $10^5$. High imaginary part
of the metallic components of the LH structure does not mean that there are
a lot of losses present, as it was assumed in Ref. \cite{Garcia_1}.

\begin{figure}[t]
\epsfig{file=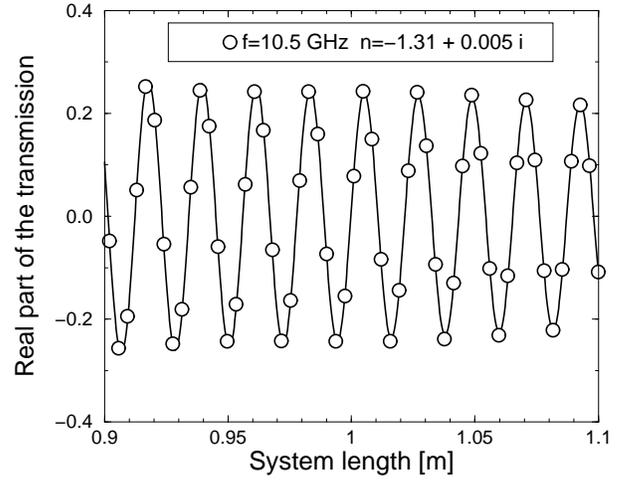,width=8cm}
\caption{System length dependence of the transmission $t$ ($T=t^*t$)  
of the EM wave of
the frequency $f= 10.5$ GHz. Symbols represent transfer matrix data, 
solid line is a
fit  $a_0e^{-\kappa x}\cos(kx+x_0)$.
Presented data correspond to the system with 
$\nne=-1.31 + 0.005~i$.
Note that system length is in meters.}
\label{0xt_5}
\end{figure}

In fig. \ref{0xt_5} we present a detailed system length 
dependence of the transmission for $f=10.5$ GHz, obtained by TM simulations. 
The length of the system was up to  300 unit cells, which corresponds 
to a system of  length equal to 1.1  m. 
From the exponential decrease of the transmission amplitude 
we  estimate the imaginary part of the refraction index  
to be  only Im $n=5\times 10^{-3}$. 

\begin{figure}[t]
\psfig{file=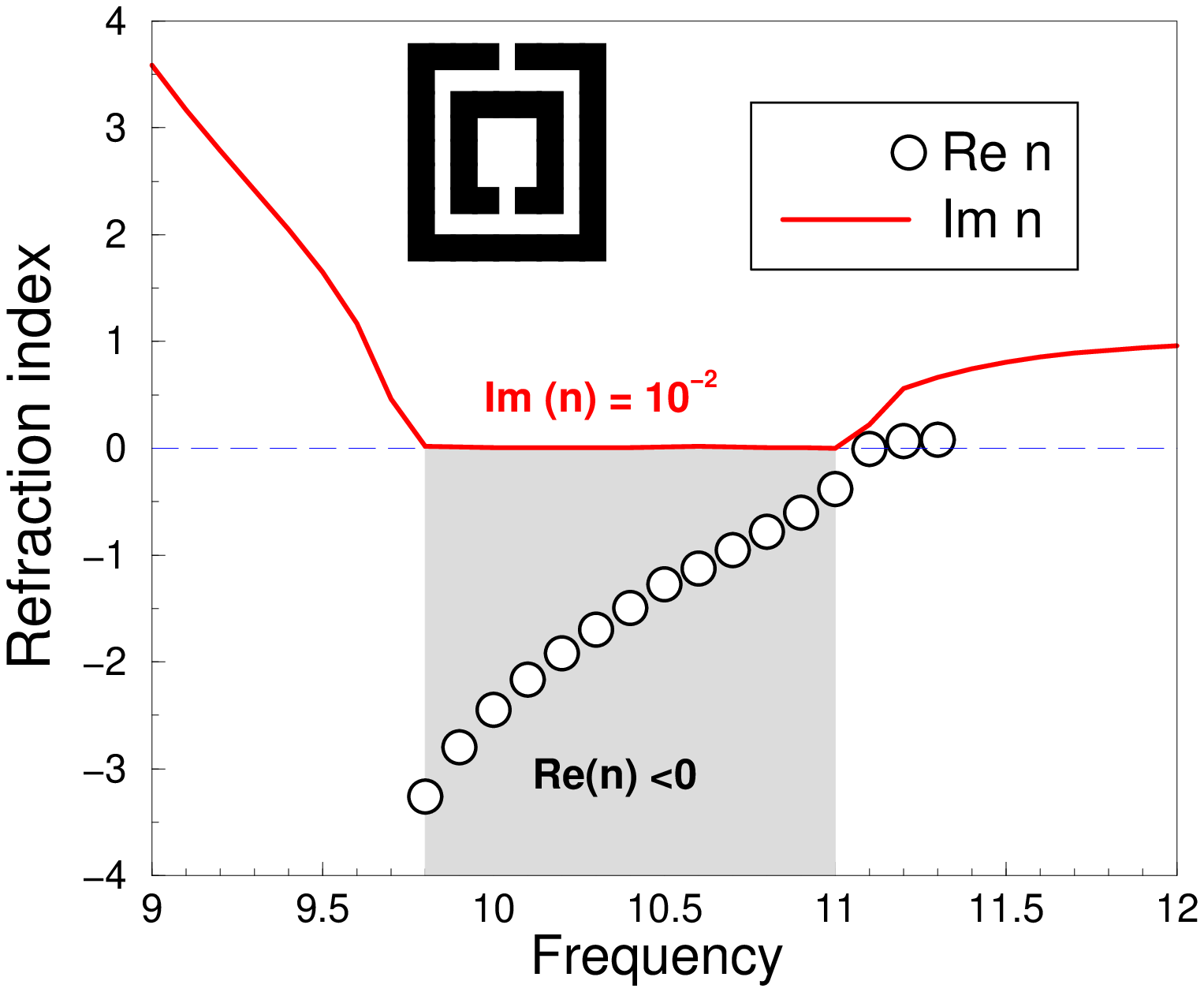,width=7cm}\\
\epsfig{file=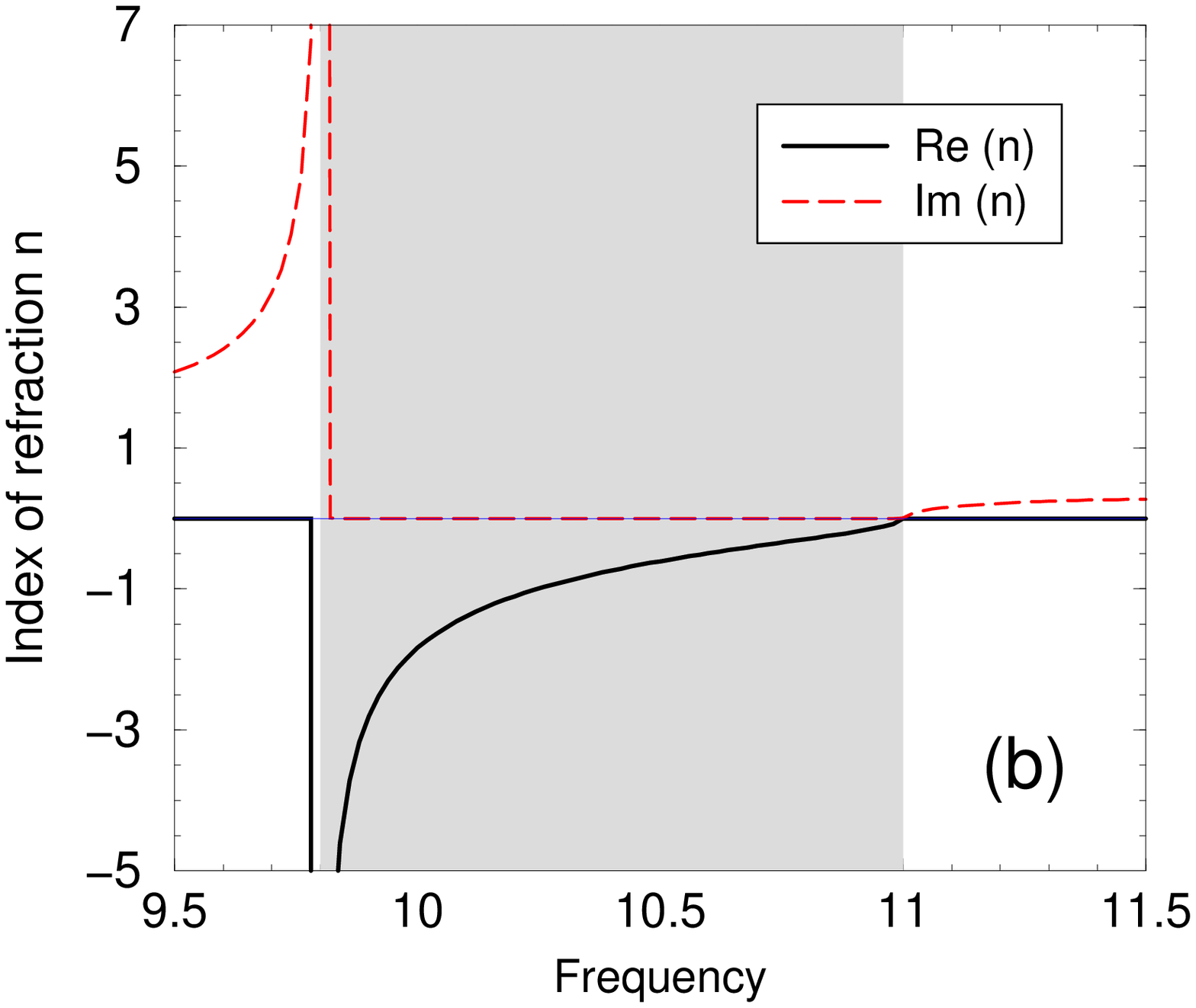,width=7cm}
\caption{Effective index of refraction (real and imaginary part) as 
a function of the frequency $f$. (a) $n$ calculated from numerical
transfer matrix  data,
(b) $n$ 
given by Eqs. \ref{epseff} and \ref{mueff} (for values of the 
material parameters
see caption of fig. \ref{0xL}.
Only negligible changes of this behavior have been observed
when $f_{e0}>0$ (data not presented).
Note that the imaginary part of  $n$ is very small. }
\label{yn}
\end{figure}

Transfer matrix data for the transmission and the reflection 
of EM waves provides us with the
complete information needed to extract  the effective parameters of the system.
Inverting the equations for the transmission and reflection of
 the {\sl homogeneous} slab of material with a given refraction
 index and impedance, we find  the refraction index.
\cite{SSMS}
We present in fig. \ref{yn} the 
effective refraction index as was obtained from the numerical data.
For comparison, we present also data for the refraction index,
calculated from the 
frequency dependent  $\epse$ and
$\mue$ given by 
Eqs. \ref{epseff} and \ref{mueff}, and 
$n=\sqrt{\epse\mue}$.
Both the numerical data and the homogeneous model
give, in  the resonance frequency interval,  negative Re $n$
with typical resonance behavior in the vicinity of the left interval edge.
We also obtain  very small imaginary part for  the refraction index.
In particular, for the wave shown in fig. \ref{0xt_5} we find that
\be
n(f=10.5~{\rm GHz})= -1.31 + i~0.005.
\ee
These parameters guarantee good transmission properties.

\medskip

\begin{figure}[t]
\epsfig{file=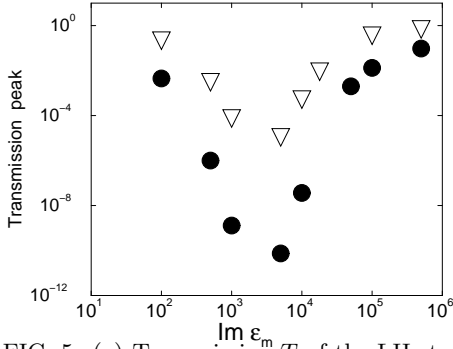,width=6cm}
\caption{(a) Transmission $T$ of the LH structure 
vs Im $\epsm$.
Both ``vertical'' (circles) and ``horizontal'' (triangles)
 orientation of SRR were
considered. \cite{note_1}
}
\label{garcia_1}
\end{figure}

It is sometimes  argued
\cite{Garcia_1}, that the high metallic permittivity of metallic
components  causes high losses in the LH structures. This is, however, 
not true. To show how the transmission of the LH structure depends
on the permittivity $\epsm$ of metallic components,  we 
have simulated LH systems
with Im $\epsm$ increasing from 0 up to $5\times 10^5$. 
As it is shown in fig. \ref{garcia_1}, an increase of the imaginary part
of the metallic permittivity improves the transmission properties
of LH materials, provided that  Im $\epsm> 10^4$. Although the 
transmission decreases to small  values
for Im $\epsm\approx 10^3-10^4$, it starts to increase and 
is of the order of one for Im $\epsm\ge 5\times 10^5$. 
As the conductance of the copper 
$\sigma$ is $5.9\times 10^7(\Omega {\rm m})^{-1}$, 
the imaginary part of the  permittivity of copper in GHz region is
of the order of $10^7$. \cite{Jackson} We expect therefore  
the transmission of a realistic
systems to be even better than the one  displayed in
fig. \ref{garcia_1}. \cite{note_1}
Our data clearly prove that the metallic components of the LH structures
can not be responsible for the high losses observed in the 
experimental studies of transmission
\cite{Smith,Shelby,Ekmel}.
As the LH systems are  highly dispersive \cite{SSMS},
and still transparent, we believe that the dispersion 
is not the  cause for the high  losses in the LH structures.

\medskip

\begin{figure}[t]
\epsfig{file=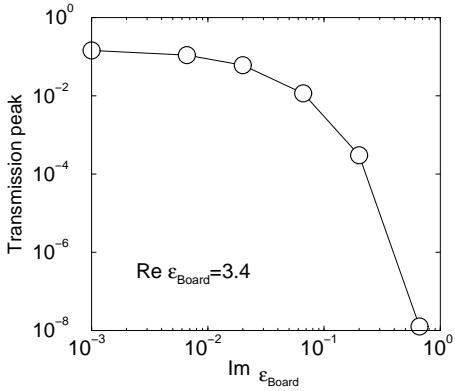,width=6cm}
\vspace*{4mm}
\caption{Transmission peak as a function of the imaginary part of the
permittivity of the dielectric board. The metallic permittivity
$\epsm=(-3+5.88~i)\times 10^{5}$. The length of the system is 10 unit
cells. Data represents the maximal transmission observed in the resonance
peak. 
}
\label{588_imag_3}
\end{figure}

To explain the relatively low transmission, observed in 
the experimental data, we have  
studied the dependence of the transmission on other
material parameters.
As the most probable mechanism of losses we consider  the 
 absorption of EM waves due to nonzero imaginary part of the  
dielectric board, on
which the metallic components are positioned.  
To test this  hypothesis, we repeated our numerical simulations for the same
structure but with a
small imaginary part to the permittivity of the dielectric board:
$\epsb=3.4+i $Im $\epsb$. Figure \ref{588_imag_3} shows how the transmission
peak decreases when imaginary part of $\epsb$ increases.
Surprisingly, the transmission strongly
 decreases with the losses in the dielectric board. 

\medskip

To conclude, we presented a detailed analysis of the numerical data
for the transmission of the electro-magnetic waves through left-handed
structures. Recovered refraction index is in agreement with the predictions of
the homogeneous model with effective parameters given by Eqs. 
\ref{epseff} and \ref{mueff}.
Numerical simulations confirmed the excellent  transmission properties of
the simulated LH  systems. We found that imaginary part of the refraction index
is only $\sim 10^{-2}$. As the value of the  imaginary part of 
the metallic permittivity
in real metals is even higher that that used in our simulations, we conclude
that metallic components of the LH structures do not represent any source of
absorption. 
Much higher losses were observed due to the absorption in the dielectric
board on which SRRs are located. 

\medskip

We thank E. N. Economou for fruitful discussions.
Ames Laboratory is operated for the U.S.Department of Energy by Iowa
State University under Contract No. W-7405-Eng-82. This work was 
supported by
the Director of Energy Research, Office of Basic Science,
DARPA and NATO grant PST.CLG.978088. P.M. thanks
Ames Laboratory for its hospitality and support and Slovak Grant Agency
for partial financial support.

\begin{thebibliography}{99}
\bibitem{note_3} Let us note  that we do not need 
$\gamma$ to be  very small. Even for relatively high
values of $\gamma=0.1$ there is still an appreciate amount of transmission
through the LH structure (data not presented here).
\bibitem{note_1} For a given metallic frequency, transmission properties
depends also on the position of the SRR in the  unit cell.
As was shown in \cite{PRE}, the transmission of 
the LH structure with gaps of the SRR oriented 
along the wires is two orders of magnitude higher than that for
LH structure with SRR ``turned'' by 
90 $^{\rm 0}$.
\end{thebibliography}
\end{document}